\documentclass[reprint,amsmath, amssymb, aps, prl, longbibliography]{revtex4-1}
\usepackage{lipsum}  
\usepackage{stackengine}
\usepackage{graphicx}
\usepackage{dcolumn}
\usepackage{bm}
\usepackage{color}

\usepackage[normalem]{ulem}

\begin{document}

\preprint{APS/123-QED}

\title{Exciton-exciton interaction in transition metal dichalcogenide monolayers and van der Waals heterostructures}
\author{Daniel Erkensten$^1$, Samuel Brem$^2$, Ermin Malic$^{1,2}$}
\affiliation{$^1$ Department of Physics, Chalmers University of Technology, Gothenburg, Sweden}
\affiliation{$^2$Department of Physics, Philipps-Universit{\"a}t, 35037 Marburg, Germany}

\date{\today}% It is always \today, today,
\begin{abstract}
Due to a strong Coulomb interaction, excitons dominate the excitation kinetics in 2D materials. While Coulomb-scattering between electrons has been well studied, the interaction of excitons is more challenging and remains to be explored. As neutral composite bosons consisting of electrons and holes, excitons show a non-trivial scattering dynamics. Here, we study on microscopic footing exciton-exciton interaction in transition-metal dichalcogenides and related van der Waals heterostructures. We demonstrate that the crucial criterion for efficient scattering is a large electron/hole mass asymmetry giving rise to internal charge inhomogeneities of excitons and emphasizing their cobosonic substructure. Furthermore, both exchange and direct exciton-exciton interactions are boosted by enhanced exciton Bohr radii. We also predict an unexpected temperature dependence that is usually associated to phonon-driven scattering and we reveal an orders of magnitude stronger interaction of interlayer excitons due to their permanent dipole moment. The developed approach can be generalized to arbitrary material systems and will help to study strongly correlated exciton systems, such as moire super lattices.\\

\end{abstract}
\maketitle
%\tableofcontents

The emergence of atomically thin 2D materials, such as graphene and monolayer transition metal dichalcogenides (TMDs), has initiated a new research field offering a platform for the investigation of many-body correlations and quantum phenomena \cite{novolrev, cao2018unconventional, wu2018hubbard, yu2017moire}. The strong Coulomb interaction in TMD monolayers gives rise to excitons, dominating the optical response, relaxation dynamics and transport characteristics \cite{review1, review2, rev3, perea2019exciton}. Excitons are neutral composite bosons (cobosons) consisting of Coulomb-bound electrons and holes. Due to a complex band structure exhibiting multiple valleys, there is a variety of different exciton species in TMDs including bright,  momentum- and spin-dark intralayer excitons \cite{dark1, dark2, darkselig} as well as spatially separated interlayer excitons in van der Waals heterostructures \cite{rivera,fogler,thygesen, paulina}.  

Previous theoretical studies have addressed the fundamental many-body processes governing the phonon-driven exciton dynamics in monolayer TMDs \cite{bremcascade, darkselig, seligline, brem2020phonon, shiqin} and van der Waals heterostructures \cite{ovesen, natmat, linzhou}. In the considered  weak-excitation regime, excitons were treated as non-interacting bosons. However, as the excitation density increases, the cobosonic nature of excitons comes to the surface and exciton-exciton scattering becomes increasingly important \cite{moodyli, xxquantumwells, kyriienko}. While scattering between electrons has been treated extensively in literature \cite{malic, jahnke}, a microscopic treatment of the Coulomb interaction of excitons as neutral cobosonic quasi-particles has proven to be challenging \cite{combescot, combescot2, katsch}. The role of direct and exchange exciton-exciton interactions in TMDs has been previously discussed, but exclusively for bright excitons interacting in a monolayer \cite{kyriienko}. In particular, exciton-exciton scattering incorporating the remarkable intervalley and interlayer excitonic landscape in 2D materials has so far remained unrevealed.  

\begin{figure}[t!]
\centering 
\includegraphics[width=0.9\linewidth]{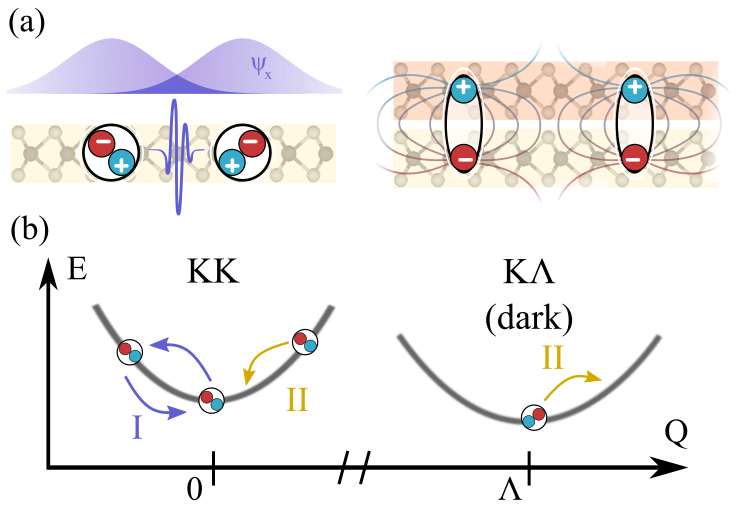}
\caption{ (a) Schematic illustration of exciton-exciton scattering in a TMD monolayer (left) and a van der Waals heterostructure (right). While the intralayer interaction is determined by the wave function overlap, the interlayer coupling resembles a dipole-dipole interaction. (b) Exemplary scattering channels in monolayer WSe$_2$ including intravalley (I, blue) and intervalley  processes (II, orange).  
}
\label{fig1}
\end{figure}

In this work, we investigate exciton-exciton scattering in TMD monolayers and van der Waals heterostructures based on a microscopic and quantum-mechanic approach. We evaluate the efficiency of the exciton-exciton interaction by calculating the excitation-induced dephasing (EID) of excitonic resonances in the incoherent limit. We resolve the underlying intra- and intervalley as well as intra- and interlayer exciton-exciton scattering channels, cf. Fig. \ref{fig1}. We reveal an intriguing temperature and screening dependence of EID and provide microscopic insights into the fundamental nature of scattering between intra- and interlayer excitons.  The latter exhibit a permanent out-of-plane dipole moment and their interaction can be considered as an efficient dipole-dipole coupling (Fig. 1(a)) resulting in an EID in the range of a few meV. In contrast, intralayer excitons do not have a permanent dipole moment and interact through higher-order electric moments induced by the internal charge inhomogeneity of excitons as cobosonic quasi-particles. Here, we show that the mass asymmetry between electrons and holes as well as the overlap of Bohr radii are the key quantities determining the exciton-exciton scattering efficiency. The gained microscopic insights are applicable to a broader class of excitonic, multi-valley materials.   \\

\begin{figure}[t!]
    \centering
    \includegraphics[width=0.8\linewidth]{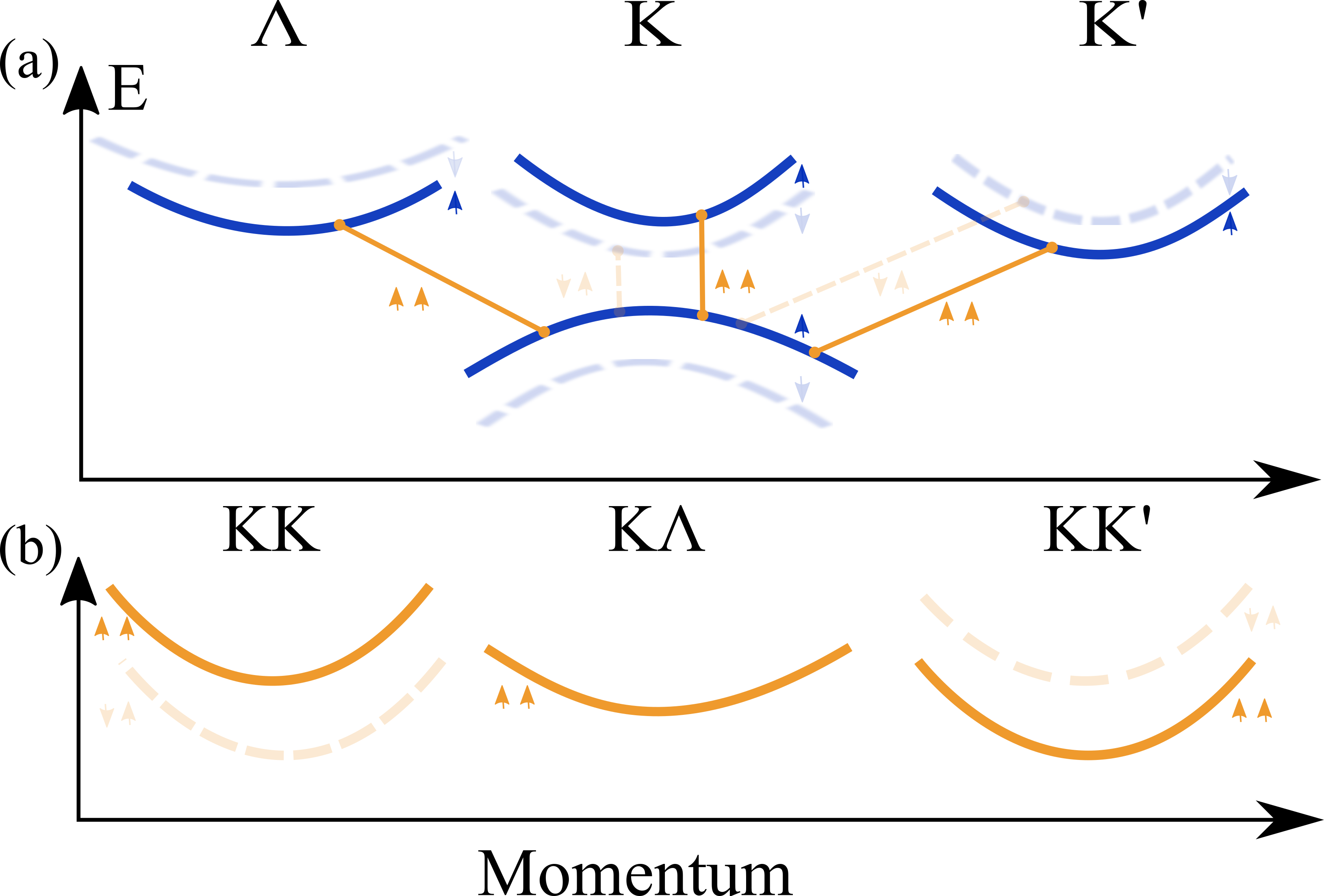} 
    \caption{ (a) Schematic illustration of the electronic band structure of monolayer WSe$_2$ in the vicinity of the high-symmmetry  K, K' and $\mathrm{\Lambda}$ points. In this work, we focus on spin-allowed transitions (yellow lines). (b) Excitonic band structures displaying the energetic ordering of  exciton states as a function of the center-of-mass momentum. Note that  the momentum-dark KK' and K$\Lambda$ states are energetically lower than the bright KK exciton. }
    \label{bandstruk}
\end{figure}

\paragraph{Theoretical model.---}To investigate exciton-exciton scattering on a microscopic footing in 2D materials, we first define the many-particle Hamilton operator. Following the approach described in Ref. \onlinecite{katsch}, we derive the excitonic Hamilton operator from the conventional electron-electron Hamiltonian by employing an identity operator expansion \cite{ivanovhaug} and rewriting the electron and hole creation and annihilation operators $c^{(\dagger)}$ and $v^{(\dagger)}$ into excitonic operators $P^{(\dagger)}$. We obtain $H=H_0+H_{x-x}$, where the interaction-free part $H_0$ includes the excitonic energies $\epsilon^{\alpha}_{\bm{Q}}=E^{\alpha}+\frac{\hbar^2 |\bm{Q}|^2}{2M^{\xi}}$, $\bm{Q}$ being the center-of-mass momentum, $M^{\xi}$ is the valley-dependent exciton mass, $E^{\alpha}$ are excitonic eigenenergies and $\alpha=(n, \xi)$. The latter is a compound index including the excitonic state $n=1s,2s...$ in the valley $\xi=(\xi_h,\xi_e)=$ KK, KK', K$\mathrm{\Lambda}$, i.e. in this work we consider the hole to be localized around the K point, while the electron lies either around the K, K' or  $\mathrm{\Lambda}$ point (also often referred to as $Q$ point in literature \cite{kormanyos}). Moreover, we focus on spin-allowed transitions (Fig. \ref{bandstruk}(a)). In principle, exciton-exciton scattering including spin-dark states is also possible, but we expect this type of scattering to be qualitatively similar to the scattering with momentum-dark states that have been considered in this work. 

The eigenenergies $E^{\alpha}$ and the associated excitonic wave functions $\varphi_{\alpha, \bm{q}}$ are obtained by solving the Wannier equation \cite{ovesen,bremcascade}, which is derived within the effective-mass approximation \cite{kirakoch}. Thus, the Wannier equation provides reliable excitonic binding energies when the exciton is well-localized in momentum space and the parabolic approximation holds \cite{eduardo2}. In the case of monolayers, we treat the screened Coulomb potential entering the Wannier equation within the Keldysh approach \cite{rytova, keldysh, cudazzo} to account for the finite width of the TMD and screening effects. For charge carriers within a heterostructure, we solve the Poisson equation for two aligned homogeneous slabs  resulting in a generalized Keldysh screening \cite{ovesen}. The obtained excitonic eigenenergies can be found in Table I in the Supplementary Material and are in good agreement both with DFT studies \cite{thygesen2} and experimental observations \cite{bindingexp}. In Fig. \ref{bandstruk}(b), the corresponding center-of-mass-dependent exciton dispersion $\epsilon^{\alpha}_{\bm{Q}}$ is schematically illustrated  for monolayer WSe$_2$ showing the $n=1s$ state of bright KK and momentum-dark KK', K$\mathrm{\Lambda}$ excitons. Since the latter are energetically lowest states, we expect that exciton-exciton scattering involving both bright and dark excitons will be efficient even at room temperature. 
\\

The second part of the Hamiltonian of interest in this work is the exciton-exciton interaction $H_{x-x}$ given by 
\begin{equation}
    H_{x-x}=\frac{1}{2}\sum_{}\mathcal{W}^{\alpha \beta \beta^{\prime}\alpha^{\prime}}_{\bm{q}}P^{\dagger}_{\alpha, \bm{Q}+\bm{q}}P^{\dagger}_{\beta, \bm{Q}'-\bm{q}}P_{\beta', \bm{Q}'}P_{\alpha', \bm{Q}} , 
    \label{xham}
\end{equation}
with $\mathcal{W}_{\bm{q}}^{\alpha \beta \beta^{\prime}\alpha^{\prime}}=(\mathcal{V}^{\alpha \beta \beta' \alpha'}_{\bm{q}}+2\mathcal{U}^{\alpha \beta \beta' \alpha'}_{\bm{q}})$ and where summation over the excitonic indices $\alpha^{(\prime)}$, $\beta^{(\prime)}$ and momenta $\bm{Q}^{(\prime)}$, $\bm{q}$ is implied \cite{katsch}. The matrix element  $\mathcal{W}_{\bm{q}}^{\alpha \beta \beta^{\prime}\alpha^{\prime}}$ includes direct electron-electron, hole-hole, and electron-hole interactions ($\mathcal{V}^{\alpha  \beta \beta^{\prime}\alpha^{\prime}}_{\bm{q}}$) and exchange interactions ($\mathcal{U}
^{\alpha \beta \beta^{\prime} \alpha^{\prime}}_{\bm{q}}$). In this work, we focus on the most relevant valley-conserving exciton-exciton scattering processes ($\alpha=\alpha^{\prime}\equiv\mu$ and $\beta=\beta^{\prime}\equiv \nu$) involving a small momentum transfer in the energetically lowest $n=1s$ state and therefore intervalley interactions have been omitted in the exciton-exciton Hamiltonian \eqref{xham}. Figure \ref{fig1}(b) illustrates exemplary scattering processes involving only KK states (process I) and including different valleys (process II). The direct part of the monolayer excitonic Coulomb matrix element reads
\begin{align}
    \mathcal{V}^{\mu \nu \nu \mu}_{\bm{q}}|_{\mathrm{mono}}=W_{\bm{q}}D_{\mu}(\bm{q})D^{*}_{\nu}(\bm{q}), \ .
    \label{interactionmono}
\end{align}
with the screened Coulomb potential $W_{\bm{q}}$ and  $D_{\mu}(\bm{q})=(F_{\mu}(\beta^{\mu} \bm{q})-F^*_{\mu}(\alpha^{\mu} \bm{q}))$ including the form factors $F_{\mu}(x\bm{q})=\sum_{\bm{q}_1}\varphi^*_{\mu, \bm{q}_1+x\bm{q}}\varphi_{\mu, \bm{q}_1}$. The latter  include the mass ratios $\alpha^{\mu}=\frac{m_e^{\mu_e}}{m_e^{\mu_e}+m_h^{\mu_h}}$, $\beta^{\mu}=1-\alpha^{\mu}$, where $m
^{\mu_{e(h)}}_{e(h)}$ is the electron (hole) mass in the valley $\mu_{e(h)}$=K, $\mathrm{\Lambda}$, K'. When considering scattering processes within a heterostructure the form factors are weighted differently depending on whether the scattering occurs between interlayer excitons exclusively or between intralayer and interlayer excitons, cf. the Supplemental Material.  The exchange contribution to the interaction reads
\begin{align}
    \mathcal{U}^{\mu\nu\nu\mu}_{\bm{q}}=\frac{1}{2}(\delta_{\mu_h, \nu_e}\hat{\mathcal{U}}^{\mu\nu}_{\bm{q}}+\delta_{\nu_h, \mu_e}\hat{\mathcal{U}}^{\nu\mu}_{-\bm{q}}) \ , 
    \label{exchange}
\end{align}
with 
$
   \hat{\mathcal{U}}^{\mu\nu}_{\bm{q}}=   \sum_{\bm{q}_1, \bm{q}_2}\tilde{W}^{\mu\nu}_{\bm{q}_2-\bm{q}_1}G_{\mu}(\bm{q}_1,\alpha^{\mu}\bm{q})G_{\nu}(\bm{q}_2, \beta^{\nu}\bm{q}).$
    Here we introduced $G_{\mu}(\bm{q}_i, x\bm{q})= \varphi^{*}_{\mu, \bm{q}_i}\varphi_{\mu, \bm{q}_i+x\bm{q}}$ with $i=1,2$ and $\tilde{W}^{\mu\nu}_{\bm{q}}=W_{\bm{q}}(\bm{q}\cdot \bm{M}_{\mu_h})(\bm{q}\cdot \bm{M}^{*}_{\nu_e})$, where $\bm{M}_{\eta}$  is the optical matrix element describing the transition probability between the conduction and valence bands in the same valley $\eta=\mu_h, \nu_e$ \cite{kirakoch, wangyaomat}. We obtain $\bm{M}_{\eta}$ using $k\cdot p$ theory \cite{luttkohn, kane} and deduce $|\bm{M}_{\eta}\cdot \bm{\mathrm{e}}_{\sigma_{\pm}}|=M_{\eta, \sigma_{\pm}}=\frac{a_0}{\sqrt{2}E_g
^{\eta \eta }}t^{\eta}(1\pm \delta_{\eta, \mathrm{K}})$ for left-handed ($\sigma_{+}$) and right-handed ($\sigma_{-}$) circularly polarized light in the vicinity of $\eta=$K. The lattice constant $a_0$, the band gap $E_g^{\eta\eta}$, and the overlap integrals $t^{\eta}= \langle u_{c,\eta}|\hat{\bm{\mathrm{p}}}|u_{v,\eta}\rangle$ with the Bloch factors $u_{\lambda, \eta}$ are extracted from DFT calculations \cite{kormanyos, wangyao}. Note that the exchange interaction is only non-zero when the hole valley of one of the two scattering excitons is identical to the electron valley of the other exciton involved in the scattering process.  
\begin{figure}[t!]
\centering 
\includegraphics[width=\linewidth]{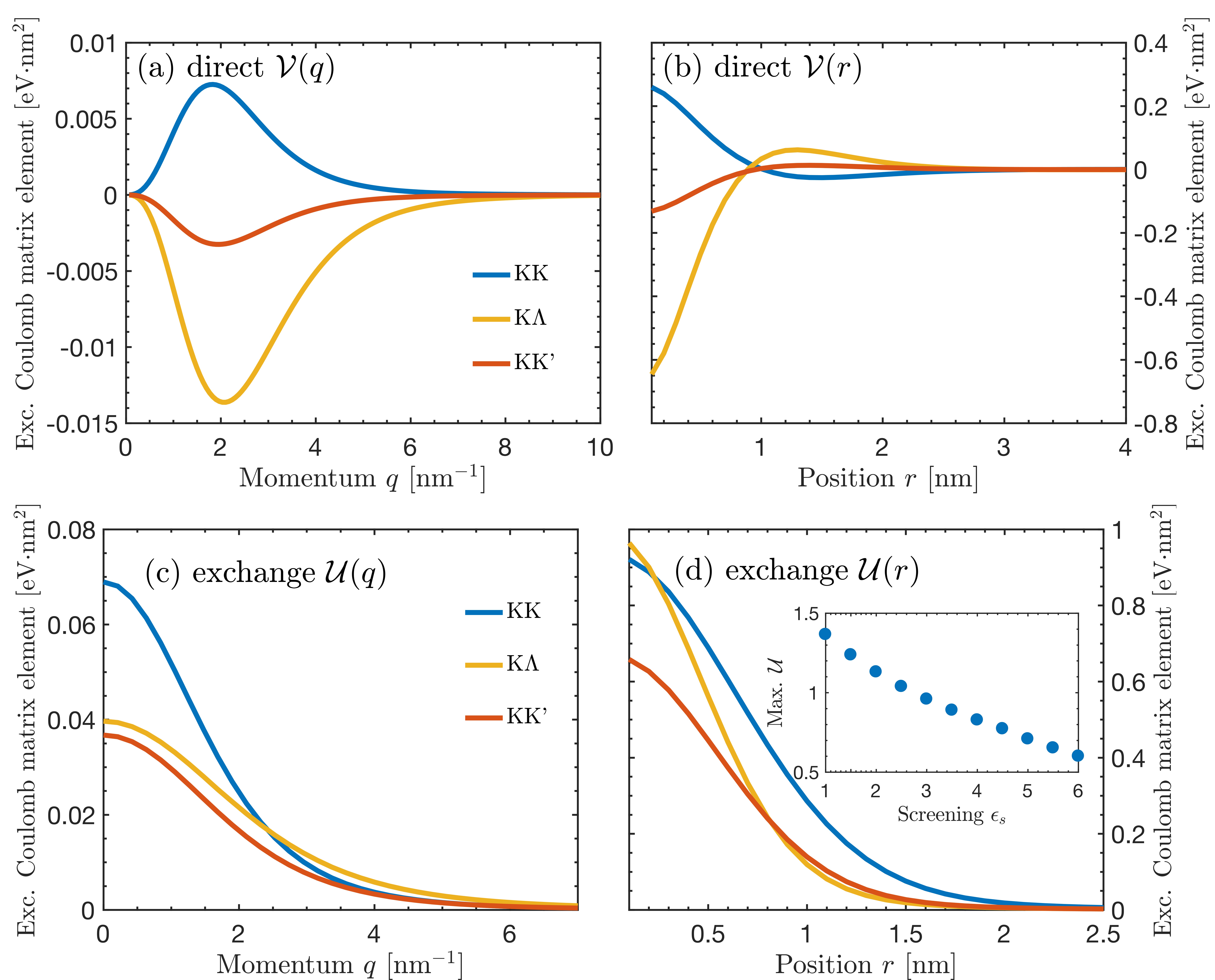}
\caption{Momentum and real space representations of the excitonic Coulomb matrix element in the WSe$_2$ monolayer. We distinguish between direct interactions $\mathcal{V}$ illustrated in (a)-(b) and exchange interactions $\mathcal{U}$ shown in (c)-(d). We consider intra- and intervalley exciton-exciton scattering (i.e. $\mu=$KK  and $\nu=$ KK, K$\mathrm{\Lambda}$, KK' respectively, cf. also Fig. \ref{fig1}(b)) in WSe$_2$. The inset in Figure (d) displays the screening dependence of the exchange interaction element with $\mu=\nu$=KK.  }
\label{matelements}
\end{figure}
Figure \ref{matelements}(a)-(d) illustrates the excitonic Coulomb matrix element in a hBN-encapsulated WSe$_2$  monolayer considering different scattering channels with $\mu=\mathrm{KK}$ and $\nu=\mathrm{KK}, \mathrm{K\Lambda}, \mathrm{KK'}$ in real and momentum space. We distinguish between direct $\mathcal{V}$ (Fig. \ref{matelements} (a)-(b)) and exchange contributions  $\mathcal{U}$ (Fig. \ref{matelements} (c)-(d)).  
We proceed by discussing these contributions separately, starting with the direct interaction. Since excitons are effectively neutral quasi-particles composed by electrons and holes, the resulting direct interaction potential in real space is reminiscent of the Lennard-Jones potential, cf. Figure \ref{matelements}(b) \cite{siee}. We find that both repulsive or attractive exciton-exciton interactions occur for different exciton species and length scales. Our model predicts that the strength of the direct interaction is determined by the mass ratios of electrons and holes of the involved excitons. This can be seen by performing a Taylor expansion of Eq. \eqref{interactionmono} for small $q$ giving $
    \mathcal{V}^{\mu\nu\nu\mu}_{q}\approx W_{q} r^2_{\mu, B} r^2_{\nu, B} q^4 Q_\mu Q_\nu$ with $r^2_{\mu, B}=\langle \mu\lvert r^2\rvert\mu\rangle$, where $r_{\mu, B}$ is the excitonic Bohr radius. Furthermore, we have introduced the effective excitonic charge 
\begin{equation}
  Q_\mu=(Q_h m_h^{\mu_h}+Q_{e} m_e^{\mu_e})/(m_h^{\mu_h}+m_e^{\mu_e})  
\end{equation}
 with $Q_{h/e}=\pm1$ determining the sign of the interexcitonic potential and thus dictating the repulsive or attractive nature of the interaction. The latter can be interpreted as a force resulting from the internal charge inhomogeneity of the exciton. In particular, an exciton with a heavy hole will be positively charged in its center, surrounded by a negatively charged shell resulting from the orbiting electron. Considering the effective excitonic charges, the matrix element in the long-range limit is always positive for $\mu=\nu$, while it becomes negative if the interacting excitons have inverted mass ratios. For example, we find an attractive character at small distances for the KK-K$\Lambda$ and KK-KK' interaction in Fig. \ref{matelements}(b). While holes are heavier than electrons for KK excitons ($Q_\mu>0$), the opposite is the case for K$\Lambda$ and KK' ($Q_\nu<0$). Moreover, having the hole at the K point and the electron at the $\mathrm{\Lambda}$ point rather than at the K'-point increases the exciton-exciton interaction by a factor of 10 in momentum space, cf. Fig.\ref{matelements}(a). This is a direct consequence of a larger mass asymmetry and consequently larger effective charge for K$\Lambda$ excitons ($m_{h}^{\mathrm{K}}=0.36m_0$ and $m_e^{\mathrm{\Lambda}}\approx{0.6 m_0}$, vs $m_{e}^{\mathrm{K}'}=0.4m_0$ \cite{kormanyos}). Moreover, the direct interaction strongly depends on the excitonic Bohr radius (expected distance between electron and hole) as it scales with the quadropole moments of the excitons.
 
 In contrast to the direct contribution, the exchange interaction $\mathcal{U}$ is non-vanishing at $\bm{q}=0$ and dominates in the considered momentum range. This behavior has already been established for interacting (bright) excitons in semiconductor quantum wells \cite{exchange1}. Moreover, the interaction is always repulsive and not highly sensitive to changes in electron and hole masses. In particular, the exchange interaction is only weakly reduced in momentum space when KK excitons are scattering with KK' excitons instead of K$\Lambda$ excitons, cf. Fig.\ref{matelements}(c). We also note that the exchange contribution can be estimated as $\mathcal{U}_{\bm{q}}\propto E_B r_B^2$, for $\bm{q}<<1$, where $E_B$ is the excitonic binding energy \cite{exchangeGlazov, kyriienko, exchangetassone}.

\begin{figure}[t!]
    \centering
    \includegraphics[width=\linewidth]{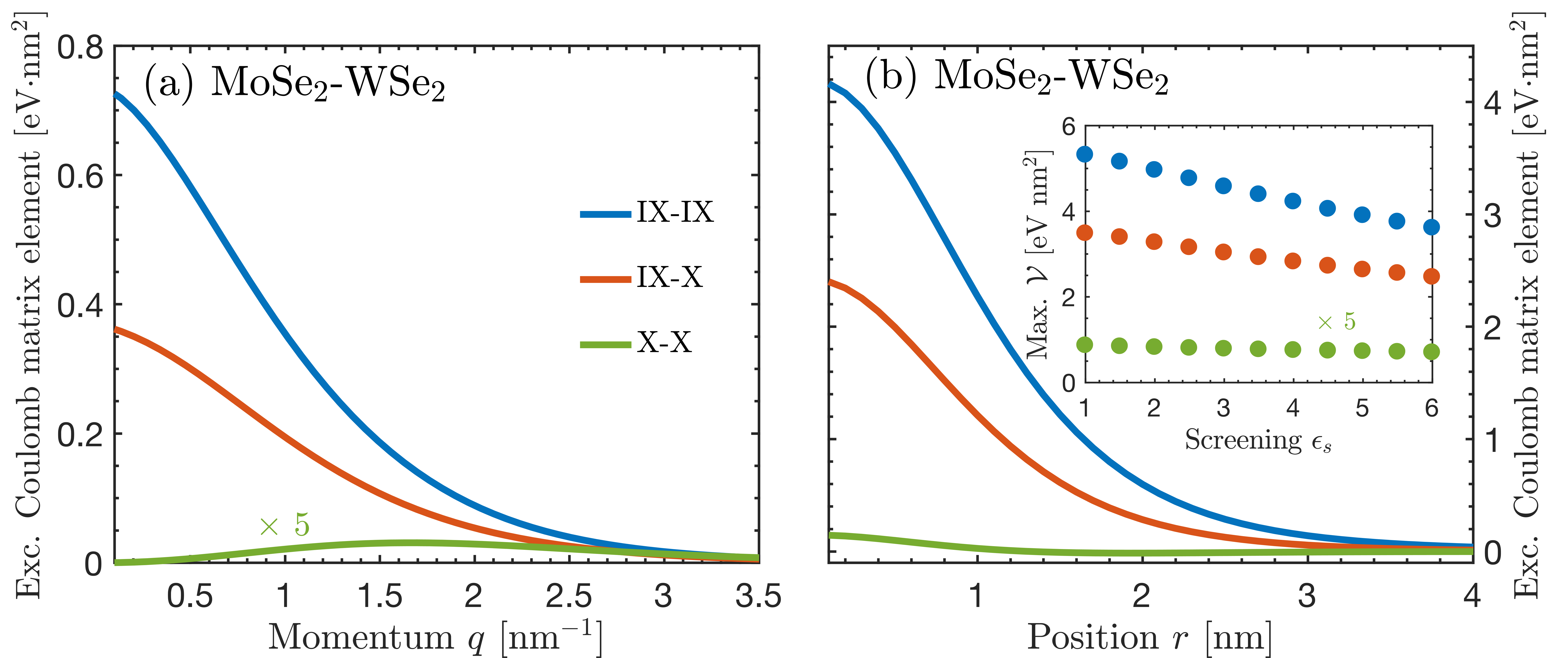}
    \caption{Momentum and real space representations of the excitonic Coulomb matrix element in the MoSe$_2$-WSe$_2$ heterostructure. We consider scattering of KK interlayer excitons with other bright interlayer excitons (IX-IX) as well as with KK intralayer excitons (IX-X) and pure intralayer scattering between KK excitons (X-X). Note that the latter is strongly suppressed. The intralayer excitons we consider here reside in the WSe$_2$ layer and the interlayer excitons are composed from holes from the MoSe$_2$ layer and electrons from the WSe$_2$ layer. The inset in Figure (b) illustrates the screening dependence of the matrix elements.}
    \label{heteroelements}
\end{figure}

Similarly to the monolayer case, we perform an analysis of the interlayer excitonic Coulomb matrix element in a van der Waals heterostructure illustrated in Figure \ref{heteroelements}(a)-(b). Due to the spatial separation of electrons and holes forming interlayer excitons, we can neglect the exchange interaction, leaving us only with the direct contribution. In fact, as shown by a recent DFT study \cite{eduardo}, the (squared) optical matrix element $|M|^2$ determining the strength of the exchange interaction is up to 200-300 times smaller for interlayer excitons in MoSe$_2$-WSe$_2$ compared to intralayer excitons.

In contrast to intralayer excitons, here the direct repulsion between electrons/holes is always stronger than the counteracting attraction of electrons and holes of different excitons. Therefore, the interaction between interlayer excitons can be understood as repulsion between two electric dipoles, cf. Fig. \ref{fig1}(a). As a consequence, the exciton-exciton interaction between interlayer excitons in the exemplary MoSe$_2$-WSe$_2$ heterostructure is up to two orders of magnitude larger than in the monolayer WSe$_2$ case, cf. Fig. \ref{matelements}(c)-(d). Importantly, the direct interlayer and intralayer matrix elements also differ qualitatively. By expanding the direct matrix elements for small momenta, we find that $\mathcal{V}
^{\mu\nu\nu\mu}_{q}|_{\mathrm{mono}}\propto W_{q}q^4$ vanishes for $q\rightarrow 0$, whereas the interlayer element remains non-zero, cf. Figs. \ref{matelements}(a) and \ref{heteroelements}(a) as well as the Supplemental Material for details.

Based on the derived excitonic Coulomb matrix elements, we now calculate the dephasing of optical polarisations induced by exciton-exciton scattering that is referred to in literature as excitation-induced dephasing (EID) \cite{haugkoch, 2020eid, newBerlin}. EID is a directly accessible phenomenon in experiments and becomes manifest as a density-dependent broadening of excitonic transitions. In contrast to another recently published work \cite{newBerlin}, we evaluate  the EID in the incoherent limit, i.e. in the long-time regime \cite{kirakoch}.  Note that we do not consider optical non-linearities, but instead focus on the linear regime that is relevant for photoluminescence and transient absorption experiments. 

We obtain microscopic access to the EID due to exciton-exciton interactions by evaluating the Heisenberg equation of motion for the excitonic polarisation $p_{\mu}=\langle P_{\mu, \bm{Q}=0}\rangle $ in the light cone, i.e. $\bm{Q}=0$. When commuting the exciton-exciton Hamiltonian $H_{x-x}$ with the polarisation, exciton-exciton correlations such as $S \propto \langle P
^{\dagger}P^{\dagger} P\rangle$ need to be considered, which previously have been treated within the Hartree-Fock approximation \cite{katsch}. Here, we go beyond and evaluate the equations of motion in second-order Born-Markov approximation \cite{haugkoch, kirakoch}, cf. the Supplemental Material. To linear order in exciton density, we obtain $ \dot{p}_{\mu}|_{H_{x-x}}=-\gamma^{\mu}_{\bm{Q}=0}p_{\mu} $ introducing the excitation-induced dephasing
\begin{equation}
    \gamma^{\mu}_{\bm{Q}}(T)=\frac{\pi}{\hbar}\sum_{\nu \bm{Q}' \bm{q}}|\mathcal{W}^{\mu\nu\nu\mu}_{\bm{q}}|^2N^{\nu}_{\bm{Q}'}(T)\delta(\Delta \epsilon) \ 
    \label{eid}
\end{equation}
with the delta-function $\Delta \epsilon=\epsilon^{\nu}_{\bm{Q'+q}}-\epsilon^{\nu}_{\bm{Q}'}-\epsilon^{\mu}_{\bm{Q+q}}+\epsilon^{\mu}_{\bm{Q}}$ ensuring the energy conservation for the considered exciton-exciton scattering processes. The appearing temperature-dependent exciton occupation  $N^{\nu}_{\bm{Q}}(T)$ is estimated with an equilibrium Boltzmann distribution, parameterized by the total exciton density $n=\sum_{\nu\bm{Q}}N
^{\nu}_{\bm{Q}}$.

In the following we focus on the EID of bright KK excitons and evaluate Eq. \eqref{eid} for the state with $\bm{Q}=0$. In general, the expression has to be evaluated numerically, but in the particular case, where intravalley scattering ($\mu=\nu=$KK) is dominant, the temperature and density dependence of EID can be addressed analytically yielding
\begin{equation}
    \gamma_0^{\mathrm{KK}}(T) \mid_{\nu=\text{KK}}=\frac{n}{\hbar}\sqrt{\frac{M^{\mathrm{KK}}}{8\pi k_B T}}\int dq |\mathcal{W}^{\mathrm{KK}}_q|^2  \ . 
    \label{EIDsimple}
    \end{equation}
We note a linear dependence of EID on exciton density $n$ as well as an explicit temperature dependence scaling with $T^{-\frac{1}{2}}$.  Normally, temperature dependencies observed in linewidth experiments are attributed to the interaction with phonons \cite{moodyli}. However, our model predicts that the temperature dependent distribution of excitons in momentum space has a direct consequence on the amount of channels available for an energy and momentum conserving exciton-exciton scattering process. This results in an additional temperature-dependent broadening which can be experimentally separated from the phonon-broadening by varying density and temperature.\\

\paragraph{Excitation-induced dephasing in
 monolayers. ---} 

\begin{figure}[t!]
\centering 
\includegraphics[width=\linewidth]{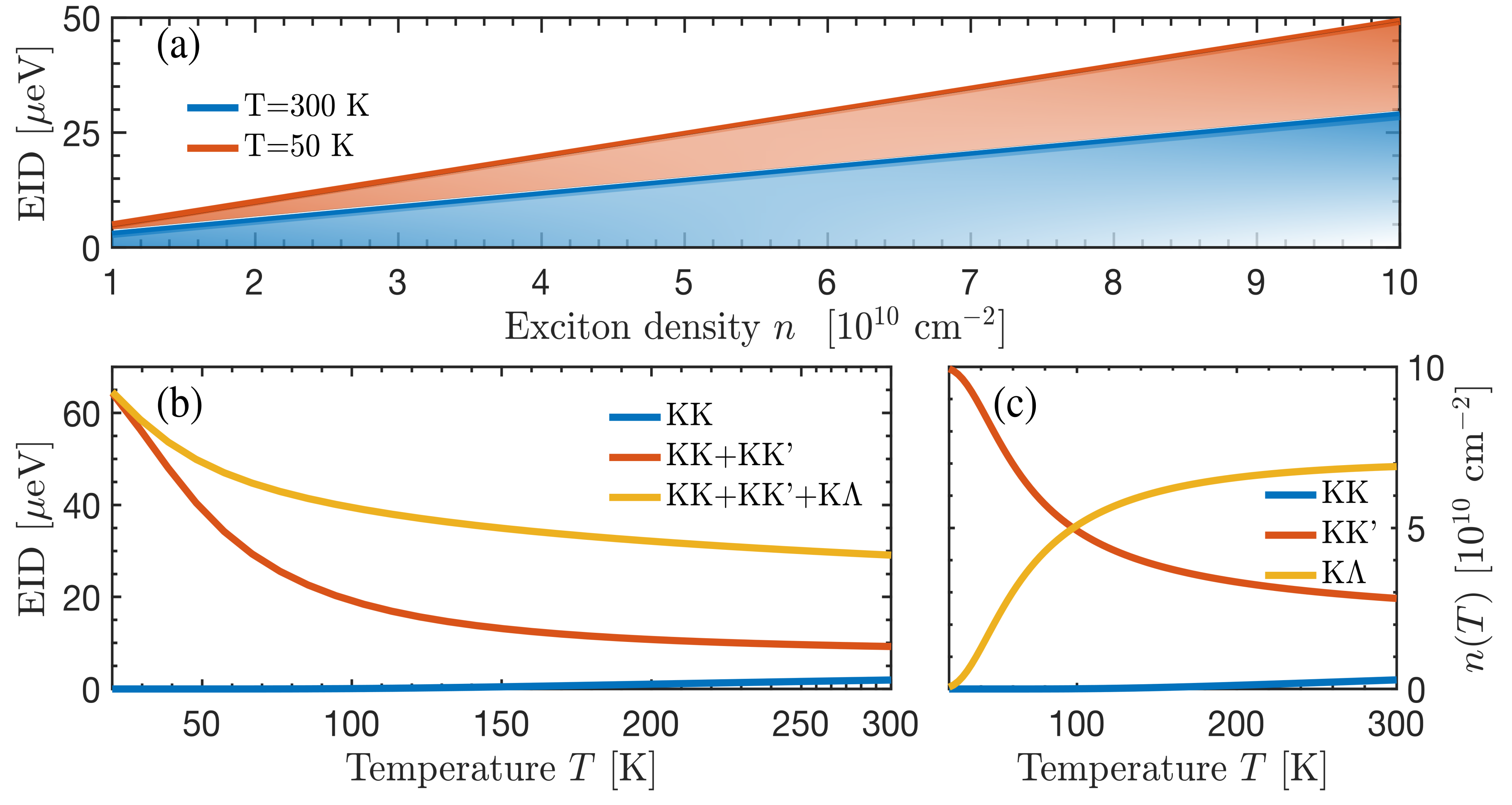}
%previous: testarmarkering2.png
\caption{Temperature and density dependence of excitation-induced dephasing in monolayer WSe$_2$. (a) EID as a function of exciton density $n$ for two different temperatures. (b) EID as a function of temperature showing single scattering contributions including intravalley scattering (KK) as well as intervalley scattering (K$\mathrm{\Lambda}$, KK') at fixed exciton density $n=10^{11} \mathrm{cm}^{-2}$. (c) Temperature-dependent density $n(T)$ of KK, K$\mathrm{\Lambda}$ and KK' excitons. }
\label{eidmono}
\end{figure}

We exploit Eqs.  \eqref{interactionmono} and \eqref{eid} to determine the excitation-induced dephasing of bright KK excitons in a hBN-encapsulated WSe$_2$ monolayer. The choice of substrate is motivated by the fact that Auger processes \cite{auger1, auger2, upconv}, not captured by our theoretical model, are shown to be suppressed in hBN-encapsulated monolayers \cite{excdiffusion}. Here, we focus on the dominant intravalley scattering channels within the K, $\Lambda$ and K' valley, cf. Fig. \ref{fig1}(b) and Fig. \ref{bandstruk}(b).
We illustrate the temperature and density dependence of the EID for the WSe$_2$ monolayer in Fig. \ref{eidmono}. As confirmed by earlier experiments \cite{moodyli, huber} and expected from our theory, the EID increases linearly with exciton density $n$, i.e. $\gamma^{\mathrm{KK}}=\gamma_{x-x}n$ with the slope $\gamma_{x-x}$. However, as demonstrated in Fig. \ref{eidmono}(b), the slope of the EID is highly temperature-dependent with $\gamma_{x-x}=2.9\cdot 10
^{-10} \ \ (4.9\cdot 10
^{-10}) \ \mu$eV$ \mathrm{cm}^2$ for $T=300$ ($50$ K). This behavior is governed by the temperature-dependence of the exciton distributions of the bright KK and momentum-dark K$\Lambda$ and KK' states (Fig. \ref{eidmono}(c)). For low temperatures ($T<30$ K),  KK' excitons determine the EID,  since most excitons reside in this energetically lowest state, see Fig. \ref{bandstruk} (b) and Table I in the Supplemental Material for the energetic hierarchy of excitonic states in WSe$_2$. At elevated temperatures the EID is dominated by K$\mathrm{\Lambda}$ excitons reflecting their highest occupation, see Fig.\ref{eidmono}(c). Hence, we have addressed the importance of dark excitonic states on the EID. Note that the intervalley scattering channels are expected to be most efficient for tungsten-based TMDs. In molybdenum-based TMDs, in particular MoSe$_2$, where the KK state is the energetically lowest state \cite{dark2}, bright excitons are expected to dominate the EID at all temperatures. 
Note that the quantitatively larger values of EID experimentally observed for the WSe$_2$ monolayer on a sapphire substrate \cite{moodyli} have been obtained from coherent nonlinear spectroscopy experiments, where incoherent contributions studied in this work cannot be accessed.

\begin{figure}[t!]
\centering 
\includegraphics[width=\linewidth]{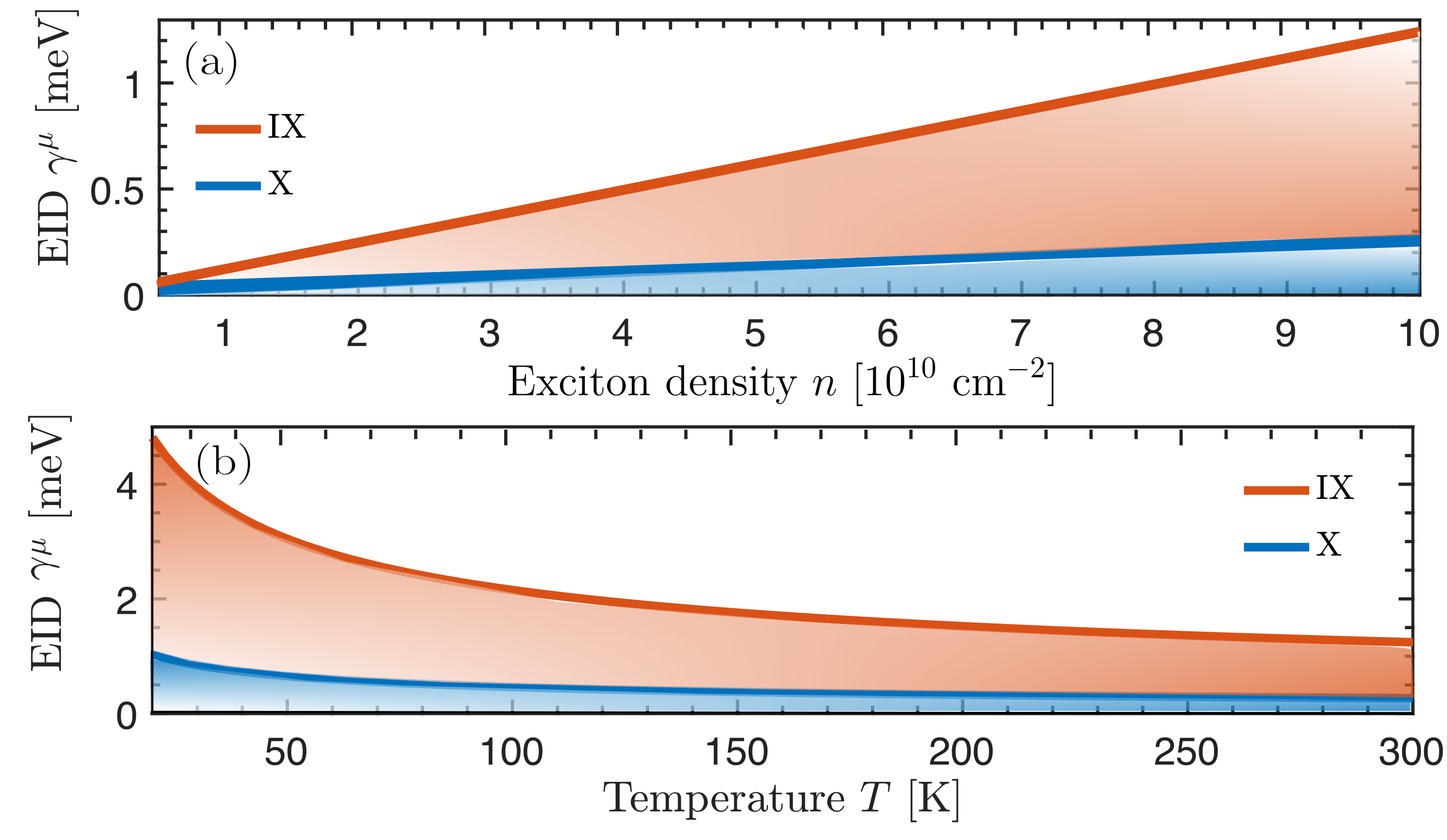}
\caption{Excitation-induced dephasing of interlayer (IX) and intralayer KK excitons (X) in the MoSe$_2$-WSe$_2$ heterostructure. (a) Density dependence of intralayer and interlayer EID at room temperature. 
(b) Temperature-dependent EID at fixed density $n=10^{11}$ $\mathrm{cm}^{-2}$ revealing  $\gamma^{\mathrm{I/IX}}(T)\propto T^{-\frac{1}{2}}$.}
\label{interlayereid}
\end{figure}

\paragraph{Excitation-induced dephasing in heterostructures. ---} 
Now, we investigate the impact of excitation-induced dephasing on a hBN-encapsulated MoSe$_2$-WSe$_2$ heterostructure. Here, both the EID of the intralayer (X) as well as of the spatially separated interlayer (IX) excitons within the heterostructure is considered. Note that we consider intralayer excitons in the WSe$_2$ layer. According to a recent DFT study \cite{interlayerdft}, KK interlayer excitons reside in the energetically lowest state in the MoSe$_2$-WSe$_2$ heterostructure. However, intervalley interlayer excitons, in particular the K$\Lambda$ state, can be found at a similar energy scale \cite{gillen}, and might also have an impact on the EID. However, treating the strong hybridisation at the $\Lambda$-point \cite{interlayerdft, bremtwist} is beyond the scope of this work and thus these states have not been considered here.
Qualitatively, the analysis of the density dependence of the intralayer and interlayer EID is similar to the monolayer case, cf. Fig. \ref{interlayereid}(a).
We find  a linear density dependence, however with drastically higher slope values of   $\gamma^{\mathrm{IX}}_{x-x}=1.2\cdot 10^{-11}$ meV$\mathrm{cm}^2$ and  $\gamma^{\mathrm{X}}_{x-x}=2.7\cdot 10^{-12}$ meV$ \mathrm{cm}^2$. Due to the stronger exciton-exciton interaction for interlayer excitons exhibiting a permanent dipole moment (Fig. \ref{matelements}), the EID is by two orders of magnitude larger compared to the  WSe$_2$ monolayer. Furthermore, due to the type-II band alignment in the MoSe$_2$-WSe$_2$ heterostructure and the large interlayer energy offset $\Delta E=315$ meV between the layers \cite{ovesen}, the interlayer exciton state is rendered by far the lowest energetic state in agreement with \cite{interlayerdft}. Note that the chosen energy offset $\Delta E$=315 meV corresponds to an AA-stacked heterostructure, but the same insights on exciton-exciton scattering are expected also for other types of stacking (AA, AA' or AB) provided a similar band offset \cite{gillen, offsetref2}.

Consequently, exclusively interlayer excitons contribute to the EID at all temperatures. In this particular case, the analytical formula in Eq. \eqref{EIDsimple} can be used to evaluate the EID predicting a $T
^{-\frac{1}{2}}$ temperature-dependence for  interlayer excitons, as observed in Fig. \ref{interlayereid}(b). 

\begin{figure}[t!]
\centering 
\includegraphics[width=\linewidth]{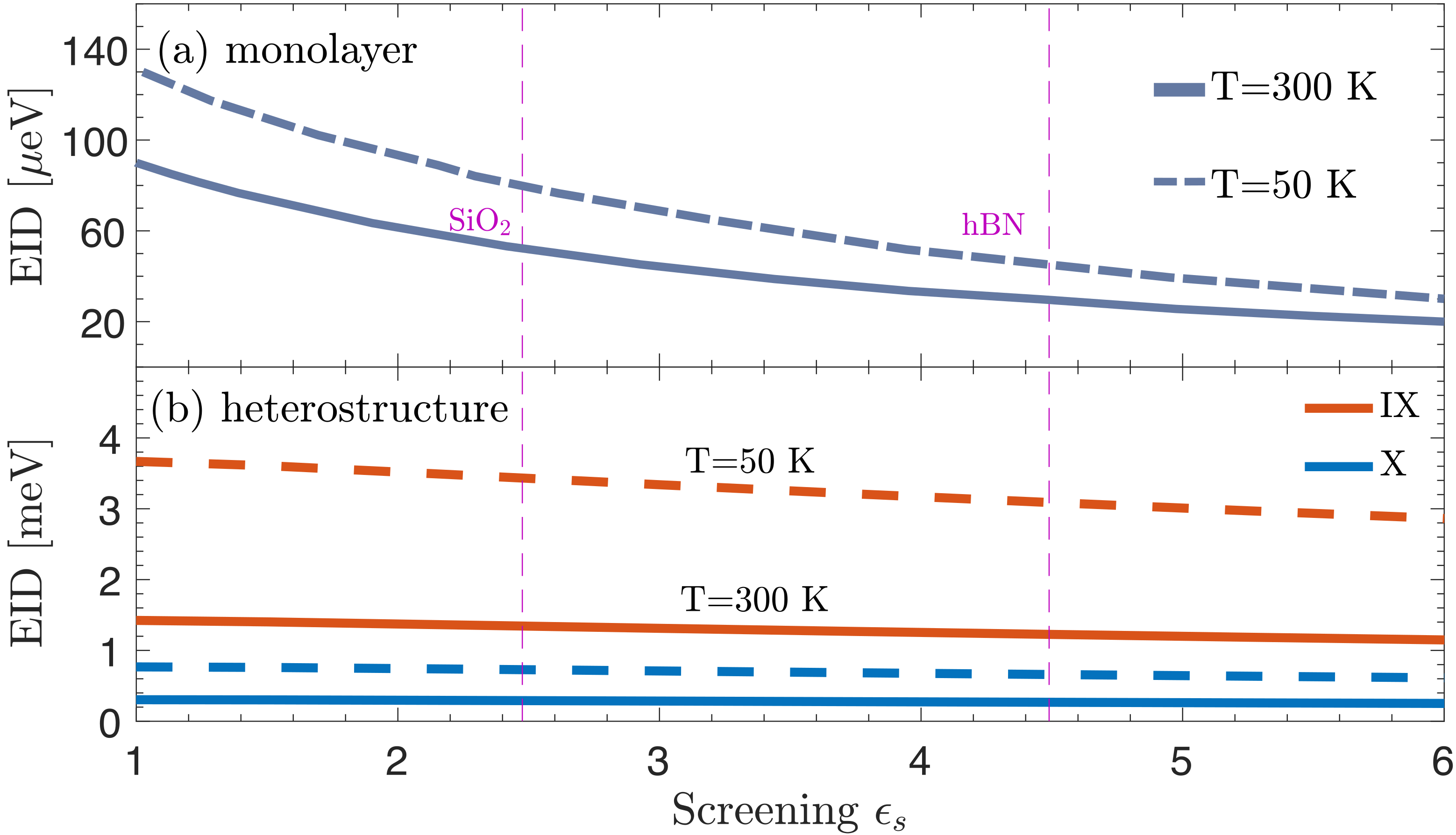}
% screeningdep.pdf
\caption{Screening dependence of excitation-induced dephasing for the (a) WSe$_2$ monolayer and the (b) MoSe$_2$-WSe$_2$ heterostructure for  $T=50$ K and $T=300$ K and at the fixed exciton density $n=10^{11}$ $\mathrm{cm}^{-2}$. Two of the most common substrates, hBN and Si$\mathrm{O}_2$, have been indicated as vertical dashed lines.}
\label{screeningstudy}
\end{figure}

Finally, we investigate how the choice of substrate affects the EID. In Fig. \ref{screeningstudy}(a)-(b) the EID is shown as a function of the background dielectric screening $\epsilon_s$ for two different temperatures. 
We find a similar behavior for $T=50$ K and $T=300$ K, namely that the EID in both the monolayer and the heterostructure displays a surprisingly moderate variation due to screening - despite the strong screening dependence of the Coulomb potential itself.    
This reflects the moderate decrease of the (direct) excitonic Coulomb matrix elements with screening, as illustrated by the insets in Fig. \ref{matelements}(d) and \ref{heteroelements}(b). The background is that a larger screening also gives rise to an enhanced Bohr radius, which increases the form factors $F_{\mu}$ and $G_{\mu}$ appearing in the excitonic Coulomb matrix elements, cf. \eqref{interactionmono} and \eqref{exchange}, respectively. Hence, there is a competition between a decreased Coulomb potential $W_{\bm{q}}$ and an increased Bohr radius resulting in a weak overall screening dependence. Still, we observe a stronger screening dependence in the monolayer case compared to heterostructures. This can be traced back to the exchange interaction that is dominating for monolayers and is more sensitive to screening than the direct contribution, which is crucial for heterostructures. In particular, the direct interaction displays a stronger dependence on the excitonic Bohr radius in the limit $\bm{q}<<1$ ($\mathcal{V}\propto r^4_{B}$) compared to the exchange interaction ($\mathcal{U}\propto E_B r^2_{B}$) and hence the compensation between decreased Coulomb potential and increased Bohr radius is weakened when the exchange interaction is dominant.

\paragraph{Conclusion.---}
We have presented a microscopic and quantum-mechanic approach on exciton-exciton scattering in 2D materials and related van der Waals heterostructures. To evaluate the scattering efficiency, we calculate the excitation-induced dephasing (EID) taking into account intra- and intervalley as well as intra- and interlayer exciton-exciton scattering channels. We predict an intriguing temperature and screening dependence and explain this by shedding light into the fundamental nature of exciton-exciton scattering. Spatially separated interlayer excitons in heterostructures exhibit a permanent dipole moment and their interaction can be considered as an efficient dipole-dipole coupling resulting in an EID of a few meV. In contrast, intralayer excitons are neutral cobosonic quasi-particles. The exciton-exciton scattering efficiency is boosted by a large electron/hole mass asymmetry and Bohr radii overlap of the excitons. The gained insights can guide future experimental studies on the impact of exciton-exciton scattering on optical properties of multi-valley 2D materials.  \\
\vspace{-0.3cm}
\paragraph{Acknowledgements.---} 
We thank Raul Perea-Causin (Chalmers) and Florian Katsch (TU Berlin) for fruitful discussions. We acknowledge funding from the European Union’s Horizon
2020 research and innovation program under grant agreement No. 881603 (Graphene
Flagship) as well as from the Swedish Research Council (VR, project number 2018-00734).
\bibliography{eidreferences} 
\end{document}

\renewcommand{\theequation}{S\arabic{equation}}
\preprint{APS/123-QED}

\title{Supplementary Material \--- Exciton-exciton interaction in transition metal dichalcogenide monolayers and van der Waals heterostructures}

\author{Daniel Erkensten$^1$, Samuel Brem$^2$, Ermin Malic}
\affiliation{Department of Physics, Chalmers University of Technology, Gothenburg, Sweden}
\affiliation{Department of Physics, Philipps-Universit{\"a}t, 35037 Marburg, Germany}

\maketitle 
\date{\today}
\subsection*{Note 1. Excitonic Coulomb matrix elements}
The starting-point of this work is the exciton-exciton Hamilton operator
\begin{equation}
 H_{x-x}=\frac{1}{2}\sum_{\mu, \nu, \bm{Q}_1, \bm{Q}_2, \bm{q}} (\mathcal{V}^{\mu \nu \nu\mu}_{\bm{q}}+2\mathcal{U}^{\mu\nu\nu\mu}_{\bm{q}})P^{\dagger}_{\mu, \bm{Q}_1+\bm{q}}P^{\dagger}_{\nu, \bm{Q}_2-\bm{q}}P_{\nu, \bm{Q}_2}P_{\mu, \bm{Q}_1} \ ,  
 \label{xham}
\end{equation} 
which is expressed in the excitonic basis and derived using the identity operator approach outlined in Ref. \cite{Katsch_2019}. Here, $P^{(\dagger)}$ are excitonic annihilation (creation) operators, $\bm{Q}_1$, $\bm{Q}_2$ are center-of-mass momenta, $\bm{q}$ is the momentum transfer, and the excitonic indices $\mu$ and $\nu$ can be interpreted as valley indices, since we restrict our calculations to scattering processes ($\mu\rightarrow \mu$, $\nu\rightarrow \nu$) in the lowest energetic $1s$ state. The matrix elements $\mathcal{V}
^{\mu\nu\nu\mu}_{\bm{q}}$ and $\mathcal{U}^{\mu\nu\nu\mu}_{\bm{q}}$ constitute direct and exchange Coulomb interactions, respectively. Here we neglected the (weak) dependence on exciton center of mass momenta of the exchange interaction.

The direct excitonic Coulomb matrix element reads  for TMD monolayers \cite{katsch}
\begin{equation}
    \mathcal{V}^{\mu \nu \nu\mu}_{\bm{q}}|_{\mathrm{mono}}=V^{cccc}_{\bm{q}}F_{\mu}(\beta^{\mu}\bm{q})F^{*}_{\nu}(\beta^{\nu}\bm{q})+V^{vvvv}_{\bm{q}}F_{\mu}(\alpha^{\mu}\bm{q})F^{*}_{\nu}(\alpha^{\nu}\bm{q}) -V^{cvvc}_{\bm{q}}\bigg(F^{*}_{\mu}(\alpha^{\mu}\bm{q})F^{*}_{\nu}(\beta^{\nu}\bm{q})+F_{\mu}(\alpha^{\mu}\bm{q})F_{\nu}(\beta^{\nu}\bm{q})\bigg) 
    \label{genexp}
\end{equation}
with the form factors $F_{\mu}(x\bm{q})=\sum_{\bm{q}_1}\varphi^{*}_{\mu, \bm{q}_1+x\bm{q}} \varphi_{\mu, \bm{q}_1}$, determined by the overlap of excitonic wave functions $\varphi_{\mu, \bm{q}}$ in valley $\mu$=KK,K$\mathrm{\Lambda}$, KK' and the electron/hole mass ratios $\alpha^{\mu}=\frac{m_e^{\mu_e}}{m_e^{\mu_e}+m_h^{\mu_h}}$, $\beta^{\mu}=1-\alpha^{\mu}$ (with $m_{e(h)}^{\mu_{e(h)}}$ being the corresponding electron (hole) masses at the $\mu$=K, $\mathrm{\Lambda}$, K' point). The exciton wave functions are obtained from solving the Wannier equation using material-specific parameters (masses and dielectric constants) taken from  Ref. \onlinecite{kormanyos}. The intraband Coulomb matrix elements $V^{cccc}_{\bm{q}}$, $V^{vvvv}_{\bm{q}}$ and $V^{cvvc}_{\bm{q}}$ describe direct  electron-electron, hole-hole and electron-hole interactions, respectively. These matrix elements can be found in their general form in Ref. \onlinecite{katsch}. For reasons of simplicity, we  consider only the long-range part of the Coulomb interaction with a small momentum transfer and thus we can approximate $ V^{cccc}_{\bm{q}}\approx{V^{vvvv}_{\bm{q}}}\approx{V^{cvvc}_{\bm{q}}}\approx{W_{\bm{q}}=\frac{V_{\bm{q}}}{\varepsilon_{\bm{q}}}} $
with $V_{\bm{q}}=\frac{e_0^2 }{2 \epsilon_0 |\bm{q}|}$ being the two-dimensional bare Coulomb potential with the non-local Keldysh-like \cite{keldysh, rytova} monolayer screening $\varepsilon_{\bm{q}}$ given by \cite{brem2020phonon}  
\begin{equation}
    \varepsilon_{\bm{q}}=\kappa_1 \tanh\bigg(\frac{1}{2}\bigg[\alpha_1 d_0 q-\mathrm{ln}\bigg(\frac{\kappa_1-\kappa_2}{\kappa_1+\kappa_2}\bigg)\bigg]\bigg) \ , 
    \label{kelds}
\end{equation}
where $d_0$ is the thickness of the TMD layer, and $\kappa_{i}=\sqrt{\epsilon^{\perp}_i \epsilon_i^{\parallel}}$, $\alpha_i=\sqrt{\epsilon_i^{\parallel}/\epsilon_i^{\perp}}$ being formed from the perpendicular and parallel components of the dielectric tensor of the considered monolayer ($i=1$) and the environment ($i=2$). The used material-specific parameters can be found in Table II. \\

Within the approximation of the intraband matrix elements introduced above, the monolayer excitonic Coulomb matrix element reduces to
\begin{equation}
    \mathcal{V}^{\mu \nu \nu\mu}_{\bm{q}}|_{\mathrm{mono}}=\frac{V_{\bm{q}}}{\varepsilon_{\bm{q}}}\bigg(F_{\mu}(\beta^{\mu}\bm{q})-F^{*}_{\mu}(\alpha^{\mu} \bm{q})\bigg)\bigg(F_{\nu}(\beta^{\nu}\bm{q})-F^{*}_{\nu}(\alpha^{\nu}\bm{q})\bigg)^{*} \ . 
    \label{vmono}
\end{equation}
We find that for $\bm{q}\rightarrow 0$, i.e. for long spatial  distances in real space, the interaction vanishes. This is reasonable since, excitons are effectively neutral objects and do not interact at large distances. Furthermore, the strength of the interaction is determined by the electron-hole mass asymmetry of the excitons involved in the scattering process indicating that the fermionic substructure of excitons play a vital role in exciton-exciton scattering.    
We can gain some more intuition for the excitonic Coulomb matrix element by expressing the difference of form factors in \eqref{vmono} in real space
$
    F_{\mu}(\beta^{\mu}\bm{q})-F^{*}_{\mu}(\alpha^{\mu} \bm{q})=\sum_{\bm{r}}\varphi^{*}_{\mu}(\bm{r})(\mathrm{e}^{i\beta^{\mu}\bm{q}\cdot \bm{r}}-\mathrm{e}^{-i\alpha^{\mu}\bm{q}\cdot \bm{r}})\varphi_{\mu}(\bm{r})  
$
and considering small momenta $q$ yielding $ F_{\mu}(\beta^{\mu}q )-F^{*}_{\mu}(\alpha^{\mu} q)\approx{-q^2 Q_{\mu}}r_{\mu, B}^2 $. Here, we introduced the (squared) excitonic Bohr radius $r^2_{\mu, B}= \langle \mu| r^2 |\mu\rangle $. Note that $\langle \mu|r|\mu\rangle=0$ since the 1s excitonic wave functions are even functions. We also defined the effective excitonic charge 
\begin{equation}
    Q_{\mu}=\frac{Q_h m^{\mu_{h}}_h +Q_e m^{\mu_{e}}_e}{m^{\mu_{h}}_h+m^{\mu_{e}}_e} \ , 
\end{equation}
with $Q_{h (e)}=\pm 1$. It can now be directly seen that the excitonic Coulomb matrix element is enhanced with increasing Bohr radii:
\begin{equation}
    \mathcal{V}^{\mu\nu\nu\mu}_{q}|_{\mathrm{mono} }\approx{W_{q} r_{\mu, B}^2 r_{\nu, B}^2q^4 Q_{\mu}Q_{\nu}} \,\,\,\,\, \text{(for small $q$)}. 
    \label{s7b}
\end{equation}
As a direct consequence, when increasing the dielectric background screening, the Coulomb potential $W_{q}=\frac{V_q}{\varepsilon_{q}}$ becomes reduced, but at the same time the excitonic Bohr radius increases (due to lower binding energies), resulting in an overall moderate screening dependence of  the excitonic Coulomb matrix element. \\

Now, we turn our attention to intralayer (X) and interlayer (IX) excitons in TMD heterostructures. This requires us to modify the dielectric screening in the excitonic Coulomb matrix elements taking into account that electrons and holes can reside in different layers. First, we consider both excitons involved in the scattering process being interlayer excitons and introduce the labels $l_1$ and $l_2$ for the two different layers. Then, we make use of the intralayer and interlayer dielectric screening functions $\varepsilon
^{l_1 l_1}_q$ and $\varepsilon^{l_1 l_2}_q$ respectively, which are given in Ref. \onlinecite{ovesen}. This enables us to deduce the corresponding excitonic Coulomb matrix element for IX excitons. Assuming $l_1\neq l_2$ we generalize \eqref{genexp} to
\begin{align}
    \mathcal{V}^{\mu\nu \nu\mu}_{\bm{q}}|_{\mathrm{IX-IX}}&=V^{c_{l_1} c_{l_1}c_{l_1}c_{l_1}}_{\bm{q}}F^{*}_{\mu}(\alpha^{\mu}\bm{q})F_{\nu}(\alpha^{\nu}\bm{q})+V^{v_{l_2} v_{l_2}v_{l_2}v_{l_2}}_{\bm{q}}F_{\mu}(\beta^{\mu}\bm{q})F^{*}_{\nu}(\beta^{\nu}\bm{q}) \nonumber \\
    & \ \ \ -V^{c_{l_1} v_{l_2} v_{l_2} c_{l_1}}_{\bm{q}}F_{\mu}(\beta^{\mu}\bm{q})F_{\nu}(\alpha^{\nu}\bm{q})-V^{c_{l_2}v_{l_1}v_{l_1}c_{l_2}}_{\bm{q}}F^{*}_{\nu}(\beta^{\nu}\bm{q})F^{*}_{\mu}(\alpha^{\mu}\bm{q}) \
\end{align}
with $
    V^{c_{l_1} c_{l_1}c_{l_1}c_{l_1}}_{\bm{q}}\approx{\frac{\tilde{V}_{\bm{q}}}{\varepsilon^{l_1 l_1}_{\bm{q}}}} \ , V^{v_{l_2} v_{l_2}v_{l_2}v_{l_2}}_{\bm{q}}\approx{\frac{\tilde{V}_{\bm{q}}}{\varepsilon^{l_2 l_2}_{\bm{q}}}} \ , V^{c_{l_1} v_{l_2}v_{l_2}c_{l_1}}_{\bm{q}}=\frac{\tilde{V}_{\bm{q}}}{\varepsilon^{l_1 l_2}_{\bm{q}}} \ $, where we introduced $\tilde{V}_{\bm{q}}=2V_{\bm{q}}$ \cite{ovesen}. Note that the products of form factors are weighted differently depending on which layer the electrons/holes reside in. As a consequence, in contrast to the monolayer case the interaction is non-zero even for $\bm{q}\rightarrow 0$ and reads 
\begin{equation}
    \mathcal{V}^{\mu \nu\nu\mu}_{\bm{q}}|_{\mathrm{IX-IX}}=\frac{d_0 e_0^2 }{\epsilon_0}\bigg(\frac{1}{\epsilon_{l_1}}+\frac{1}{\epsilon_{l_2}}\bigg) \,\,\,\,\, \text{(for small $q$)} 
\end{equation}
with $d_0$ being the layer thickness (here assumed to be equal for both layers), and $\epsilon_{l_{i}}$, $i=1,2$ denoting the dielectric constants for the individual layers. We can now  also consider interlayer-intralayer exciton scattering yielding
\begin{align}
    \mathcal{V}^{\mu\nu \nu\mu}_{\bm{q}}|_{\mathrm{IX-X}}&=V^{c_{l_1} c_{l_1}c_{l_1}c_{l_1}}_{\bm{q}}F^{*}_{\mu}(\alpha^{\mu}\bm{q})F_{\nu}(\alpha^{\nu}\bm{q})+V^{v_{l_1} v_{l_2}v_{l_2}v_{l_1}}_{\bm{q}}F_{\mu}(\beta^{\mu}\bm{q})F^{*}_{\nu}(\beta^{\nu}\bm{q}) \nonumber \\
    & \ \ \ -V^{c_{l_1} v_{l_2} v_{l_2} c_{l_1}}_{\bm{q}}F_{\mu}(\beta^{\mu}\bm{q})F_{\nu}(\alpha^{\nu}\bm{q})-V^{c_{l_1}v_{l_2}v_{l_2}c_{l_1}}_{\bm{q}}F^{*}_{\nu}(\beta^{\nu}\bm{q})F^{*}_{\mu}(\alpha^{\mu}\bm{q}) \ . 
\end{align}
In the limiting case of small  $\bm{q}$ we find
 \begin{equation}
 \mathcal{V}^{\mu\nu \nu\mu}_{\bm{q}}|_{\mathrm{IX- X}}\approx{\frac{d_0e_0^2}{2\epsilon_0}\bigg(\frac{1}{\epsilon_{l_1}}+\frac{1}{\epsilon_{l_2}}\bigg)} \,\,\,\,\, \text{(for small $q$)}.
 \end{equation} 
 As performed in detail in the monolayer case, we may also expand the interlayer/intralayer excitonic Coulomb matrix elements in terms of Bohr radii,  leading to the same conclusions as above, namely that the matrix elements are enhanced by increased Bohr radii. However, in contrast to the monolayer case, the matrix elements do not scale with $q$ for small momenta, but exhibit a constant value as a consequence of the permanent dipole moment of interlayer excitons. 
 
 Now, we turn our attention to the exchange contribution of the exciton-exciton interaction. This interaction is only present, when the hole valley of one of the scattering excitons is identical to the electron valley of the other exciton involved in the scattering process. Assuming a weak dependence on exciton center of mass momenta, it reads for monolayers
 \begin{equation}
     \mathcal{U}^{\mu\nu\nu\mu}_{\bm{q}} =\frac{1}{2}\sum_{\bm{q}_1, \bm{q}_2}V^{cvcv}_{\bm{q}_2-\bm{q}_1}(G_{\mu}(\bm{q}_1, \alpha^{\mu}\bm{q})G_{\nu}(\bm{q}_1, \beta^{\nu}\bm{q})\delta_{\mu_h, \nu_e}+G^{*}_{\nu}(\bm{q}_1, \alpha^{\nu}\bm{q})G^{*}_{\mu}(\bm{q}_1, \beta^{\mu}\bm{q})\delta_{\nu_h, \mu_e}) \ , 
 \end{equation}
 with $G_{\mu}(\bm{q}_i, x\bm{q})=\varphi^{*}_{\mu, \bm{q}_i}\varphi_{\mu, \bm{q}_i+x\bm{q}} $ ($i=1,2$) and $V^{cvcv}_{\bm{q}}$ being the interband Coulomb matrix element. Considering the long-range part of the interaction, we find $V^{cvcv}_{\bm{q}}\approx{W_{\bm{q}} (\bm{q}\cdot \bm{M}_{\eta})(\bm{q}\cdot \bm{M}^{*}_{\eta})}$ with the optical matrix element $\bm{M}_{\eta}=\frac{\hbar}{m_0 E_g^{\eta\eta}}\langle u_{c, \eta}|\hat{\bm{\mathrm{p}}}|u_{v,\eta}\rangle$, including the single-particle band gap $E_g
^{\eta\eta}$, free electron mass $m_0$ and Bloch factors $u_{\lambda, \eta}$ ($\lambda=c,v$). In the vicinity of $\eta=$K the optical matrix element is projected on the left-handed ($\sigma_+$) or right-handed ($\sigma_-$) Jones vector $\bm{\mathrm{e}}_{\sigma_{\pm}}$ giving
\begin{equation}
    |\bm{M}_{\eta}\cdot \bm{\mathrm{e}}_{\sigma_{\pm}}|=\frac{a_0 t}{\sqrt{2}E^{\eta\eta}_g} (1\pm \delta_{\eta, \mathrm{K}})\ , 
\end{equation}
with the lattice constant $a_0$ and next-neighbor hopping integral $t$. These parameters are obtained from Refs. \cite{kormanyos} and \cite{wangyao} respectively. Due to the spatial separation of electrons and holes of interlayer excitons the exchange interaction in heterostructures is expected to be strongly suppressed and has therefore been neglected. \\

\subsection*{Note 2. Excitation-induced dephasing} 
We  start from the simplified exciton-exciton Hamiltonian in \eqref{xham} and determine the equation of motion for the excitonic polarisation $p_{\mu, \bm{Q}}=\langle P^{\dagger}_{\mu, \bm{Q}}\rangle$. The evolution of time-dependent quantum-mechanical operators is governed by the Heisenberg equation of motion $i\hbar \partial_t \langle \cdot \rangle =\langle [\cdot, H]\rangle $ \cite{kirakoch} resulting in
\begin{equation}
    -i\hbar \partial_t p_{\mu, \bm{Q}} =\epsilon^{\mu}_{\bm{Q}}p_{\mu, \bm{Q}} +\sum_{\nu \bm{Q}_1 \bm{q}}(\mathcal{V}^{\mu  \nu \nu \mu}_{\bm{q}}+2\mathcal{U}^{\mu\nu\nu\mu}_{\bm{q}})\underbrace{\langle P^{\dagger}_{\mu, \bm{Q+q}}P^{\dagger}_{\nu, \bm{Q}_1-\bm{q}}P_{\nu, \bm{Q}_1} \rangle^{c}}_{:= S^{\mu \nu \nu}_{\bm{Q+q}, \bm{Q}_1-\bm{q}, \bm{Q}_1}} \ . 
    \label{peq}
\end{equation}
Here we only consider the correlated part $\langle  P^{\dagger}_{\mu, \bm{Q+q}}P^{\dagger}_{\nu, \bm{Q}_1-\bm{q}}P_{\nu, \bm{Q}_1} \rangle^{c}$ and neglect single-particle factorizations. It is also important to note that we treat the excitonic operators $P^{(\dagger)}$ as bosonic operators, i.e. $[P_1, P^{\dagger}_2]=\delta_{1,2}$. In the high excitation regime, the fermionic substructure of excitons has to be taken into account when evaluating the equations of motion and \emph{cobosonic} commutator relations have to be used \cite{katsch}. However, our Hamiltonian is constructed in such a way that it includes fermionic correction terms compensating for the fact that we use the bosonic commutator relations to derive the equations of motion. For instance, it can be shown that the difference between applying the bosonic commutator and a cobosonic commutator 
to the direct electron-electron and hole-hole contributions becomes manifest only in different prefactors \cite{katsch}.
 The direct electron-hole scattering part of the excitonic Hamiltonian was verified by explicitly evaluating the equation of motion for the polarisation in the electron-hole picture and transforming the result into the excitonic picture up to the second-order in excitonic operators and comparing with the corresponding result obtained using the excitonic Hamiltonian. The second-order exchange term ($\mathcal{U}$) is also added to the Hamiltonian as a cobosonic correction term. When deriving the excitonic Hamiltonian using an identity operator expansion \cite{ivanovhaug} of the conventional carrier-carrier Hamiltonian only a first-order exchange interaction $H_{\mathrm{exch}}\sim \hat{\mathcal{U}}P^{\dagger}P$ is obtained \cite{katsch}. The second -rder exchange term ($\propto P^{\dagger}P^{\dagger}PP$) is found by commuting $\hat{\mathcal{U}}$ with the excitonic polarisation in a bosonic and a cobosonic way and comparing the results. Hence, we construct and add the second-order exchange interaction to the Hamiltonian in such a way that the fermionic correction terms induced by the cobosonic commutator are captured on the level of the Hamiltonian. The first-order term $\hat{\mathcal{U}}$ results in a small energy renormalization ($\sim$ 20 meV \cite{Katsch_2019}) for KK excitons and is neglected in this work.

Since we want to go beyond the Hartree-Fock approximation in this work, we derive an equation of motion for the appearing new quantity $S^{\mu \nu \nu}_{\bm{Q+q}, \bm{Q}_1-\bm{q}, \bm{Q}_1}$ in \eqref{peq} yielding 
\begin{align}
    i\hbar \dot{S}^{\mu \nu \nu}_{\bm{Q+q}, \bm{Q}_1-\bm{q}, \bm{Q}_1}&=
    \sum_{\sigma}\bigg(\mathcal{W}^{\nu\sigma \nu\mu}_{\bm{Q}_1-\bm{Q}-\bm{q}}+\mathcal{W}^{\nu\sigma \mu\nu}_{\bm{q}}\bigg) \bigg(N^{\nu}_{\bm{Q}_1-\bm{q}}N^{\mu}_{\bm{Q+q}}-N^{\nu}_{\bm{Q}_1-\bm{q}}-N^{\nu}_{\bm{Q}_1-\bm{q}}N^{\nu}_{\bm{Q}_1}-N^{\mu}_{\bm{Q+q}}N^{\nu}_{\bm{Q}_1}\bigg)p_{\sigma, \bm{Q}} \\ \nonumber 
    & \ \ \ \ \ +\Delta \epsilon S^{\mu \nu \nu}_{\bm{Q+q}, \bm{Q}_1-\bm{q}, \bm{Q}_1}     \label{eqS}
\end{align}
with $\mathcal{W}^{\mu\nu \rho\sigma}_{\bm{q}}=\mathcal{V}^{\mu\nu\rho\sigma}_{\bm{q}}+2\mathcal{U}^{\mu\nu\rho\sigma}_{\bm{q}}$ and  $\Delta \epsilon=( \epsilon^{\nu}_{\bm{Q}_1}-\epsilon^{\nu}_{\bm{Q}_1-\bm{q}}-\epsilon^{\mu}_{\bm{Q}+\bm{q}})$. Here, we  performed correlation expansions of occurring three-operator quantities ($\propto\langle  P^{\dagger}P^{\dagger}P\rangle $) and five-operator quantities ($\propto \langle P^{\dagger}P^{\dagger}P^{\dagger}PP\rangle $), employing the random-phase approximation (RPA) \cite{kirakoch,haugkoch} and we neglected higher-order correlations (corresponding to the Born approximation). For instance, the five-operator quantity $\langle P^{\dagger}_{a_1}P^{\dagger}_{a_2}P^{\dagger}_{a_3}P_{a_4}P_{a_5}\rangle $ ($a_i$, $i=1...5$ being treated as a compound index) can be disentangled by using the following cluster expansion: 
\begin{align}
   \langle P^{\dagger}_{a_1}P^{\dagger}_{a_2}P^{\dagger}_{a_3}P_{a_4}P_{a_5}\rangle &\approx{\langle P^{\dagger}_{a_1}\rangle \langle P^{\dagger}_{a_2}P_{a_4}\rangle^c }\langle P^{\dagger}_{a_3}P_{a_5}\rangle^c + \langle P^{\dagger}_{a_1}\rangle \langle P^{\dagger}_{a_2}P_{a_5}\rangle^c \langle P^{\dagger}_{a_3}P_{a_4}\rangle^c \nonumber \\
   & \ \ \ \langle P^{\dagger}_{a_2}\rangle \langle P^{\dagger}_{a_1}P_{a_4}\rangle^c \langle P^{\dagger}_{a_3}P_{a_5}\rangle^c +\langle P^{\dagger}_{a_2}\rangle \langle P^{\dagger}_{a_1}P_{a_5}\rangle^c \langle P^{\dagger}_{a_3}P_{a_4}\rangle^c \\
   & \ \ \ \langle P^{\dagger}_{a_3}\rangle \langle P^{\dagger}_{a_1}P_{a_4}\rangle^c \langle P^{\dagger}_{a_2}P_{a_5}\rangle^c+\langle P^{\dagger}_{a_3}\rangle \langle P^{\dagger}_{a_1}P_{a_5}\rangle^c \langle P^{\dagger}_{a_2}P_{a_4}\rangle^c +\langle P^{\dagger}_{a_1}P^{\dagger}_{a_2}P^{\dagger}_{a_3}P_{a_4}P_{a_5}\rangle^{c} \nonumber \ ,
\end{align}
with the correlated part $\langle P^{\dagger}_{a_1}P^{\dagger}_{a_2}P^{\dagger}_{a_3}P_{a_4}P_{a_5}\rangle^{c}$. Note that this type of factorization procedure only holds if the excitons are treated as pure bosons.
Here, we only kept terms proportional to $\langle P^{\dagger}_{a_i}\rangle $, which  enables us within RPA to find an equation of motion coupling back directly to the excitonic polarisation. Within the RPA, $\langle P^{\dagger}P\rangle^c\rightarrow N$ with $N$ being the exciton occupation. The resulting equation of motion in \eqref{eqS} can be solved within the Markov approximation \cite{kirakoch} yielding
\begin{align}
    S^{\mu \nu \nu }_{\bm{Q+q}, \bm{Q}_1-\bm{q}, \bm{Q}_1}&=-i\pi\sum_{\sigma}( \mathcal{W}^{\nu\sigma \nu\mu}_{\bm{Q}_1-\bm{Q}-\bm{q}}+\mathcal{W}^{\nu\sigma \mu\nu}_{\bm{q}})(N^{\nu}_{\bm{Q}_1-\bm{q}}N^{\mu}_{\bm{Q+q}}-N^{\nu}_{\bm{Q}_1-\bm{q}}-N^{\nu}_{\bm{Q}_1-\bm{q}}N^{\nu}_{\bm{Q}_1}-N^{\mu}_{\bm{Q+q}}N^{\nu}_{\bm{Q}_1})p_{\sigma, \bm{Q}}\delta(\Delta \epsilon)  
    \label{smu}
\end{align}
with $\Delta \epsilon= \epsilon^{\nu}_{\bm{Q}_1}-\epsilon^{\nu}_{\bm{Q}_1-\bm{q}}+\epsilon^{\sigma}_{\bm{Q}} -\epsilon^{\mu}_{\bm{Q+q}}$.  Substituting \eqref{smu} into \eqref{peq}, only taking the diagonal (resonant) part $\sigma=\mu$, and keeping only the direct scattering terms ($\propto \mathcal{V}_{\bm{q}}\mathcal{U}_{\bm{q}}, |\mathcal{V}_{\bm{q}}|^2,|\mathcal{U}_{\bm{q}}|^2$) and the linear term in exciton occupation enables us to arrive at a  simple equation of motion
\begin{equation}
    \partial_t p_{\mu, \bm{Q}}|_{H_{x-x}}=-\gamma^{\mu}_{\bm{Q}}  p_{\mu, \bm{Q}} \ , 
\end{equation}
where the excitation-induced dephasing is  apparent reading
\begin{equation}
    \gamma^{\mu}_{\bm{Q}}=\frac{\pi}{\hbar}\sum_{\nu \bm{Q}_1 \bm{q}}|\mathcal{V}^{\mu \nu \nu \mu}_{\bm{q}}+2\mathcal{U}^{\mu \nu \nu \mu}_{\bm{q}}|^2 N^{\nu}_{\bm{Q}_1}\delta\big(\epsilon^{\nu}_{\bm{Q}_1+\bm{q}}-\epsilon^{\nu}_{\bm{Q}_1}+\epsilon^{\mu}_{\bm{Q}}-\epsilon^{\mu}_{\bm{Q+q}}\big) \ . 
    \label{eid}
    \end{equation}
Here we used  the relations $\mathcal{V}^{\mu\nu\nu\mu}_{\bm{q}}=\mathcal{V}^{\nu\mu \mu \nu}_{-\bm{q}}=(\mathcal{V}^{\nu\mu \mu \nu}_{\bm{q}})^{*}$ (similarly for $\mathcal{U}$) and shifted indices $\bm{Q}_1\rightarrow \bm{Q}_1+\bm{q}$.

\subsection*{Note 3. Energies, masses, Bohr radii and material-specific parameters}
In Table I, we provide the excitonic eigenenergies and masses which are required to evaluate the the excitation-induced dephasing in hBN-encapsulated WSe$_2$ monolayers and the MoSe$_2$-WSe$_2$ heterostructure. The excitonic energies are expressed relative to the KK-state, i.e. given by $E^{\mu}=
\Delta_{\mu, \mathrm{KK}}-E^{\mu}_{b}$, with $\Delta_{\mu, \mathrm{KK}}\equiv E^{\mu}_{\mathrm{single}}-E^{\mathrm{KK}}_{\mathrm{single}}$, $E^{\mu}_{\mathrm{single}}$ being the single-particle separation between the top of the valence band in valley K and bottom of the conduction band dispersion relation in valley K, K' and $\mathrm{\Lambda}$ for $\mu=\mathrm{KK, KK', K\Lambda}$, and $E^{\mu}_b$ is the excitonic binding energy extracted from the Wannier equation \cite{haugkoch}. Note that we explicitly take into account screening from the TMD layer(s) and the environment when evaluating the binding energies with the Keldysh-like screening function being given by \eqref{kelds} (see the Supplemental Material of \cite{ovesen} for the generalization to heterostructures). The necessary parameters to evaluate the corresponding screening functions are found in Table II. We find a good agreement between the obtained binding energies and recent DFT studies \cite{thygesen2, gillen} and experiments \cite{bindingexp}. For a direct comparison with free-standing monolayer WSe$_2$ and heterostructure MoSe$_2$-WSe$_2$ we refer to our previous work, \cite{brem2020phonon} and \cite{ovesen} respectively.  

Due to the large interlayer energy offset $\Delta E\approx{315}$ meV (corresponding to a AA-stacked heterostructure \cite{gillen}) in MoSe$_2$-WSe$_2$ , KK is rendered by far the lowest energetic state \cite{ovesen}.  
We also include the corresponding Bohr radii, which enter directly the direct excitonic Coulomb matrix elements expanded around small momenta, cf. Eq. \eqref{s7b}. The obtained radii are in sound agreement with recent experiments \cite{bindingexp}.  
\newpage 
\begin{table}[h!]
\begin{tabular}{|l|l|l|l|l|l|}
\hline 
\rule{0pt}{10pt} Valley $\mu=(\mu_h, \mu_e)$ & $\Delta_{\mu, \mathrm{KK}}$ (meV) &

$E
^{\mu} $ (meV) & $r_{\mu, B}$ (nm) & $m^{\mu_h}_h$ ($m_0$)      & $m^{\mu_e}_e$ ($m_0$)   \\ \hline
KK         & - & -169 (-117)  & 1.72                 & 0.36  & 0.29   \\ \cline{1-6}
K$\mathrm{\Lambda}$ & -5&  -213 (-134)       & 1.31                & 0.36                  & 0.6   \\ \cline{1-6}
KK'         & -37  & -224 (-179) & 1.50                & 0.36                  & 0.4   \\ \hline
KK (interl. MoSe$_2$-WSe$_2$) &- & -407 & 1.59 & 0.6 & 0.29 \\ \hline 
\end{tabular}
\caption{Effective hole mass $m^{\mu_h}_h$,  electron masses $m^{\mu_e}_e$, Bohr radius $r_{\mu, B}$, excitonic energy $E^{\mu}$, and single-particle energy separations $\Delta_{\mu, \mathrm{KK}}$ for different intra- and intervalley exciton species in hBN-encapsulated monolayer WSe$_2$. Note that two numbers are given in the third column for intralayer energies. The latter refers to the excitonic energy of intralayer WSe$_2$ excitons in the MoSe$_2$-WSe$_2$ heterostructure, which are seen to be reduced due to the increased screening from the heterostructure. The last line in the table also includes data for KK MoSe$_2$-WSe$_2$ interlayer excitons. Note that $E^{\mathrm{KK-interl}}$ includes the interlayer energy separation $\Delta E=315$ meV.  The excitonic energies and wave functions are obtained from solving the Wannier equation. The effective masses are expressed as fractions of the free electron mass, $m_0$, and are, as the single particle separations, taken from Ref.  \onlinecite{kormanyos}. The Bohr radius is computed as the average distance between the electron and the hole, i.e. $r_{\mu, B}=\sqrt{\langle \mu| r^2 |\mu \rangle}$, with the expectation value taken with respect to the 1s exciton wave functions in the corresponding valley $\mu=\mathrm{KK}, \mathrm{K}$$\mathrm{\Lambda}, \mathrm{KK'}$. }
\end{table}
\begin{table}[h!]
    \centering
  \begin{tabular}{|c|c|c|}
\hline 
    Parameter  & - & Ref.\\ \hline 
    $d_0$ (nm) & 0.67 (0.65) & \cite{dielectricref} \\ \hline 
  $\epsilon^{\perp}_{\mathrm{TMD}}$  & 7.5 (7.4) & \cite{dielectricref}  \\ \hline 
   $\epsilon^{\parallel}_{\mathrm{TMD}}$  & 15.1 (13.3) & \cite{dielectricref} \\ \hline 
    $\kappa_{\mathrm{hBN}}$ & 4.5 & \cite{hbnref} \\ \hline 
\end{tabular}
    \caption{Material-specific parameters used to evaluate the screened Coulomb potential in monolayer WSe$_2$ and heterostructure MoSe$_2$-WSe$_2$ (data for MoSe$_2$ in parenthesis). }
    \label{tab:my_label}
\end{table}
\bibliography{supplrefs}